\documentstyle[preprint,aps,epsf]{revtex}

\begin{document}
\draft
\tightenlines

\title{
Quadratic diffusion Monte Carlo and pure estimators for atoms
}


\author{A. Sarsa~$^1$, J. Boronat~$^2$, and J. Casulleras~$^2$}
\address{$^1$ International School for Advanced Studies, SISSA,\\
Via Beirut 2/4, I-34014, Trieste, Italy \\
$^2$ Departament de F\'\i sica i Enginyeria Nuclear, Campus Nord B4-B5, \\ 
Universitat Polit\`ecnica de Catalunya, E-08034 Barcelona, Spain}

\date{\today}
\maketitle

\begin{abstract}
The implementation and reliability of a quadratic diffusion Monte Carlo
method for the study of ground-state properties of atoms are discussed. We
show in the simple yet non-trivial calculation of the binding energy
of the Li atom that the method presented is effectively second-order
in the time step.
The fulfilment of the expected quadratic behavior  relies on some basic
requirements of the trial wave function used for importance sampling, in
the context of the fixed-node approximation. Expectation values of radial
operators are calculated by means of a pure estimation based on the forward
walking methodology. It is shown that accurate results without
extrapolation errors can be obtained with a pure algorithm, explicitely
reported, that can be easily implemented in any previous diffusion 
Monte Carlo program.

\end{abstract}


\pagebreak

\section{Introduction}
\label{intro}

Quantum Monte Carlo (QMC) methods have been widely employed in quantum
chemistry calculations \cite{anderson,arnow,reynolds1,hammond}. Nowadays,
QMC methods   have achieved accuracies comparable with CI calculations,
mainly for small atoms and molecules. QMC is specially well suited to
obtain the ground-state energy, providing an accurate treatment of
electronic correlations in atomic and molecular systems. Apart from some
technical and/or algorithmic differences, all the QMC methods pursue a
common objective which is to solve the Schr\"odinger equation by  relating
the Green's function of the system to a random walk.  In the present work,
we will focus on the diffusion Monte Carlo (DMC) method,  based on a short
time approximation to the Green's function. The required antisymmetry of
the atomic and molecular ground-state wave functions  leads to the
well-known sign problem. The present analysis is restricted to the
standard fixed-node (FN) approximation \cite{anderson,reynolds1}. According to the
FN method, a prefixed nodal  surface is defined in the configuration space
of the system. That model nodal surface, which does not change along the
calculation, introduces a boundary condition that forbids any crossing
through a node. It has been proved that the resulting FN energy  is an
upper bound to the exact ground-state energy \cite{reynolds1},  the
quality of the upper bound being related to the overlap between the model
and the exact nodal surfaces.  

The DMC method is simpler than domain Green's function Monte Carlo
\cite{gfmc} but, contrary to this, DMC presents a time-step $\Delta t$
dependence that requires additional analysis. In general, the use of a
finite $\Delta t$  to  approximate the imaginary-time Green's function
introduces a bias in the calculation. That problem is addressed by using
several small time steps and then  extrapolating  the results for the
energy to the $\Delta t \rightarrow 0$ limit. The simplest implementation
of DMC uses a short-time Green's function which is accurate to order
$(\Delta t)^2$ (Eq. \ref{dmc.eq8l}). In this case, one obtains energies that behave linearly with 
sufficiently small $\Delta t$ and then an extrapolation to zero time-step 
is unavoidable. In
order to reduce that large time-step dependence, and hence accelerate the
rate of convergence of the calculation, several algorithms have been
proposed  \cite{vrbik,rothstein,chin,umrigar,mella}. Reynolds {\em et al.}
\cite{reynolds1,hammond}  introduced an accept/reject step after the
gaussian and drift movements  in order to exactly sample $(\Phi_0)^2$ 
for all time steps if the exact  wave function $\Phi_0$ 
were used for
importance sampling. Since in a true calculation the trial wave function is
close to the exact one,  the time-step dependence is significantly
decreased with respect to the linear case. On the other hand, the rejection
method introduces the problematic issue of persistent configurations
\cite{bueckert,umrigar}, 
giving rise to a somehow unpredictable behaviour
of the final $\Delta t$ dependence of the energy.

A second alternative to reduce the time-step bias is to work out
second-order algorithms by considering short-time Green's function accurate
to order $(\Delta t)^3$ (Eq. \ref{dmc.eq8}). Some time ago, Chin \cite{chin} developed second-order
propagators  and proved the quadratic efficiency of them in several model
problems. That approach, here referred to as quadratic DMC (QDMC) method, has
been profusely used in the microscopic study of quantum fluids, mainly
$^4$He \cite{boro4he} and $^3$He \cite{boro3he}. In all these applications, QDMC has proved to be a
real second-order algorithm and therefore the time-step bias is much more
under control than in linear DMC algorithms. It is worth  noticing that
liquid $^3$He is a fermion system, and the presence of a fixed nodal surface
(FN approximation) has not broken down the expected second-order accuracy
\cite{boro3he}.
This feature contradicts some preliminary conclusions \cite{umrigar} 
that questioned a 
$(\Delta t)^2$ behavior when using nodal trial functions. Recently, Forbert
and Chin \cite{forbert} have improved the time-step accuracy by developing a fourth-order
algorithm that has manifested a fourth-order power law in a exigent calculation
like the binding energy of liquid $^4$He.

The use of importance sampling ($\psi$) in the DMC solution of the Schr\"odinger
equation makes walkers to be distributed according to the mixed 
probability distribution $\psi \Phi_0$, and not $(\Phi_0)^2$. Therefore, mixed
estimators, which are the natural output in DMC, are only
unbiased when the operators in question coincide with the Hamiltonian $H$ 
or commute with it. In a
first and very common approximation the bias of mixed estimators of,
for instance, radial operators are dealt with an extrapolation method
\cite{gfmc}.
Within this procedure, a less biased estimation is obtained from a
proper linear combination of variational and mixed estimators. The
efficiency of that linear extrapolation is strongly related to the
variational quality of $\psi$, and the results so obtained are still biased
in a quantity difficult to assess \cite{east,casulleras1}. It is possible, 
however, to eliminate 
the uncertainties present in the extrapolation method using really pure
estimators. 

The objective of obtaining pure estimators has been accomplished  by
using different techniques mainly bilinear sampling \cite{bilinear}, path
integral formalism \cite{baroni,sarsa1} and forward walking 
\cite{reynolds2,east,barnett1,runge,casulleras1,langfelder}. Among them, 
several implementations of forward walking have been the most explored
algorithms to achieve unbiased estimations of radial operators. Following 
Liu {\em et al.} \cite{liu}, the necessary ingredient to correct the mixed
estimator, i.e., the  quotient $(\Phi_0({\bf R})/\, \psi({\bf R}))$, can
be obtained from the asymptotic offspring of the ${\bf R}$ walker. In some
implementations of that forward walking procedure, tagging algorithms
\cite{reynolds2,barnett1,runge} have been devised in order to estimate the
asymptotic number of descendants. The complexity of an efficient tagging
algorithm, together with the large fluctuations observed in the asymptotic
offspring, have impeded its widespread use among the DMC community. 
However, there is a simpler algorithm to
sample pure estimators which does not require any tagging algorithmic
structure and that can be very easily incorporated in any DMC code 
\cite{casulleras1}. This method has
been checked in model problems and has proved its accuracy in many
calculations of quantum liquids properties.

In this work, we address the two above discussed aspects of the DMC method: 
the problem of the time-step dependence and 
the sampling of the exact distribution to extract pure estimators. 
We propose
the use of a second-order DMC algorithm, that presents small time-step
errors, and  unbiased estimators to calculate 
exact values for non differential properties. As it is 
commented in the
next section, the implementation 
of that algorithm can be done rather straightforwardly in any DMC code. 
The reliability of the method has been analyzed  
by studying the ground
state of atomic lithium. This simple but not trivial system has been chosen 
since,  on one hand,  presents all the ingredients to test the performance
of the method, and on the other, there are very accurate calculations 
\cite{king1,yan,king2,galvez1} with which we can compare our results.

The rest of the paper is organized as follows. In Sec. \ref{method}, the
QDMC algorithm and the pure estimation methodology is presented. 
Sec \ref{results} is devoted to present and discuss the
results. Finally, the main conclusions of the present work are presented in 
Sec. \ref{conclusions}. 

\section{Method}
\label{method}

\subsection{Quadratic Diffusion Monte Carlo}

The DMC method solves the Schr\"odinger equation in imaginary time
for the function $f({\bf R},t)=\psi({\bf R}) \Phi({\bf R},t)$,
\begin{equation}
-\frac{\partial
f({\bf R},t)}{\partial t}= -\frac{1}{2} \, \mbox{\boldmath $\nabla$}_
{{\bf R}}^2 f({\bf R},t) +
\frac{1}{2} \, \mbox{\boldmath $\nabla$}_{{\bf R}}( {\bf F}({\bf R}) 
f({\bf R},t) )
 +(E_L({\bf R})-E) \,
f({\bf R},t) \ ,
\label{dmc}
\end{equation}
$\Phi({\bf R},t)$ being the wave function of the system and
$\psi({\bf R})$ a trial function used for importance sampling. Equation
(\ref{dmc}) is written in atomic units, which are the ones used throughout
this work. In
Eq.  (\ref{dmc}), $E_L=\psi({\bf
R})^{-1} H \psi({\bf R})$ is the local energy and ${\bf F}({\bf
R})=2
\psi({\bf R})^{-1} \mbox{\boldmath $\nabla$}_{\bf R} \psi({\bf R})$;
${\bf R}$ stands for a
$3N$-coordinate vector and $E$ is an arbitrary energy shift.  

The r.h.s. of Eq. (\ref{dmc}) may be written as the action of three
operators $O_i$ acting on the wave function $f({\bf R},t)$,
\begin{equation}
-\frac{\partial f({\bf R},t)}{\partial t} = (O_1+O_2+O_3)\,
f({\bf R},t) \equiv O\, f({\bf R},t)
\label{dmc.eq4p}
\end{equation}
The three terms $O_i$ are interpreted by similarity with classical
differential equations. The first one, $O_1$, corresponds to a free
diffusion with a diffusion coefficient $D=1/2$; $O_2$ acts as a driving force
due to an external potential, and  finally $O_3$ looks like a birth/death
term. In Monte Carlo, the Schr\"odinger equation (\ref{dmc.eq4p}) is best
suited when it is written in a integral form by introducing the Green's
function $G({\bf R}^{\prime},{\bf R},\Delta t)$, which gives the
transition probability  from an initial state ${\bf R}$ to a final one
${\bf R}^{\prime}$ in a time $\Delta t$, 
\begin{equation}
     f({\bf R}^{\prime},t+\Delta t) =\int G({\bf R}^{\prime},{\bf R},
\Delta t)\, f({\bf R},t)\, d{\bf R} \ .
\label{dmc.eq6}
\end{equation}
More explicitly, the Green's function is given in terms of the operator $O$ by
\begin{equation}
    G({\bf R}^{\prime},{\bf R}, \Delta t) =
    \left \langle\,
{\bf R}^{\prime}\, | \, \exp(-O \Delta t)\, |\, {\bf R}\, \right \rangle.
\label{dmc.eq7}
\end{equation}                  

Most applications of the DMC method work with  the simplest version of the
short-time approximation,
\begin{equation}
\exp \left( -O \Delta t \right) = \exp \left( -O_3 \Delta t \right)
\exp \left( -O_2 \Delta t \right) \exp \left( -O_1 \Delta t \right) +
{\cal O} \left(
(\Delta t)^2 \right)  \ .
\label{dmc.eq8l}
\end{equation}
This expansion generates a linear time-step dependence that is
reproduced in any calculation when the dependence of the energy on $\Delta
t$ is analyzed. A significantly  better behavior is obtained by expanding
the exponential of the operator $O$  to higher orders in $\Delta t$. A good
compromise between algorithmic complexity and efficiency is obtained by
using a second-order expansion (QDMC). In this case, the
Green's function $G({\bf R}^{\prime},{\bf R},\Delta t)$ is approximated
by                                                       
\begin{eqnarray}
 \exp \left( -O \Delta t \right) & = &
     \exp \left( -O_3 \frac{\Delta
t}{2} \right ) \, \exp \left( -O_2 \frac{\Delta t}{2} \right ) \,
\exp \left( -O_1 \Delta t \right ) \label{dmc.eq8} \\[0.6cm]
& & \times \ \exp \left( -O_2 \frac{\Delta t}{2} \right ) \,
\exp \left( -O_3 \frac{\Delta t}{2} \right )  + {\cal O} \left( (\Delta t)^3
\right)
\, . \nonumber
\end{eqnarray}
This decomposition, which is the one we have used in our calculations,
is not unique as pointed out by Rothstein {\em et al.} \cite{rothstein} and
Chin \cite{chin}.
Introducing the above expansion (6) in Eq. (\ref{dmc.eq6}) the
Schr\"odinger equation, written in integral form, becomes 
\begin{eqnarray}
f({\bf R}^{\prime},t+\Delta t) & = & \int \left[
G_3 \left ({\bf R}^{\prime},{\bf R}_1,\frac{\Delta t}{2} \right )
G_2 \left ({\bf R}_1,{\bf R}_2,\frac{\Delta t}{2} \right )
G_1 \left ({\bf R}_2,{\bf R}_3,\Delta t \right ) \right .  \\[0.6cm]
 &  &  \left .
\times \ G_2 \left ({\bf R}_3,{\bf R}_4,\frac{\Delta t}{2} \right )
G_3 \left ({\bf R}_4,{\bf R},\frac{\Delta t}{2} \right )  \right] \,
f({\bf R},t)\, d{\bf R}_1 \ldots d{\bf R}_4 d{\bf R} \ . \nonumber
\label{dmc.eq9}
\end{eqnarray}    

In Eq. (7), the total Green's function $G$ is split into the
product of individual Green's functions $G_i$, each one associated to the
single operator $O_i$. $G_1$ is the Green's function corresponding to the
free diffusion term ($O_1$), and thus it is the well-known solution for a
noninteracting system,                                                          \begin{equation}
G_1({\bf R}^{\prime},{\bf R}, t)  =  (4 \pi D
t)^{-\frac{3N}{2}}\, \exp \left[ - \frac{({\bf R}^{\prime}-{\bf R})^2}{4 D
t} \right] \ .
\label{dmc.eq10p}
\end{equation}
In the MC simulation, the evolution given by $G_1$ corresponds to
an isotropic gaussian movement of size proportional to $\sqrt{D t}$. The
Green's function $G_2$ describes the movement due to the drift force
appearing in $O_2$; its form is given by \cite{vrbik}
\begin{equation}
G_2({\bf R}^{\prime},{\bf R},t) =  \delta \left(
{\bf R}^{\prime}-{\bf R}(t) \right), \ \ \ \ \ \ \mbox{where} \left \{
\begin{array}{l}
{\bf R}(0)={\bf R}  \\[0.4cm]
\frac{\displaystyle d {\bf R}(t)}{\displaystyle d t}=D\, {\bf
F}({\bf R}(t)) , \end{array}
\right. \ .                                                    
\label{dmc.eq10s}
\end{equation}
Under the action of $G_2$, the walkers  evolve in a deterministic way
according to the drift force ${\bf F}({\bf R}(t))$. In order to preserve the
second-order accuracy in the time step, the differential equation
(\ref{dmc.eq10s}) must  also be solved with a second-order integration
method. Finally, the third individual Green's function $G_3$ has an
exponential form with an argument that depends on the difference between the local
energy of a given walker and the prefixed value $E$,
\begin{equation}
G_3({\bf R}^{\prime},{\bf R}, t) = \exp \left[ -(E_L({\bf R})-E)\, t  
\right ]\, \delta({\bf R}^{\prime}-{\bf R}) .
\label{dmc.eq11}
\end{equation}
This third term, which is called the branching factor, assigns a weight to each
walker according to its local energy. Depending on the value of this weight
the walker is replicated or eliminated in the population list. A pseudo
code
which describes  the evolution of a walker for a time step $\Delta t$ 
is presented in Ref. \cite{AppendixA}. It is worth noticing that a second-order
algorithm, like the one presented here, does not introduce any new terms
with respect to a linear approximation. Therefore, it is quite simple to
move from a linear code to a second-order one from the programmer's point
of view. 

As mentioned earlier, instead of using higher-order evolution operators 
some authors have been
using a linear DMC method which incorporates an accept/reject step (Metropolis DMC
-MDMC-) \cite{reynolds1,hammond}. After a proposed gaussian plus drift movement for an individual
electron, the new position is accepted with probability
\begin{equation}
p = {\rm min} \left( \frac{|\psi({\bf R}^\prime)|^2 \, G({\bf R},{\bf
R}^\prime,\Delta t)}{|\psi({\bf R)}|^2 \, G({\bf R}^\prime,
{\bf R},\Delta t)} , \, 1 \right) \ .
\label{metro1}
\end{equation}
The detailed balance condition (\ref{metro1})
ensures theoretically a more accurate sampling of $|\psi|^2$, 
but introduces  in DMC the potential problem of persistent
configurations.  Results obtained with the MDMC method show
a significant reduction of the time-step bias, with respect to DMC, but its
efficiency depends on the particular system under study and the
extrapolation law to $\Delta t=0$ can show erratic behaviors
\cite{forbertt}.

\subsection{Pure Estimators}

When the asymptotic limit ($t \rightarrow \infty$) is reached, the sampling
of an operator $O$ is carried out according to the mixed distribution
$\psi \Phi_0$, with $\Phi_0$ the ground-state wave function. Thus, the natural
output in DMC corresponds to the so called {\em mixed} estimators. The
mixed estimator of an operator $O({\bf R})$ is, in general,  biased by the trial
 wave
function $\psi$ used for importance sampling. Only when $O({\bf R})$ is 
either the Hamiltonian, or
commutes with it, the mixed estimation is the
exact one. A simple method that has been widely used to 
approximately remove the bias present in the mixed estimations 
 is the extrapolated estimator \cite{gfmc},
\begin{equation}
\langle O({\bf R}) \rangle_e=2 \, \langle O({\bf R})
\rangle_m - \langle O({\bf R}) \rangle _v \ ,
\label{dmc.extrap}
\end{equation}                            
from the knowledge of the mixed estimator $\langle O({\bf R})\rangle_m$ and
the variational one
\begin{equation}
\langle O({\bf R}) \rangle_v=
\frac{\langle \psi({\bf R}) \, | \, O({\bf R}) \, | \,
\psi({\bf R}) \rangle}
{\langle \psi({\bf R}) \, | \, \psi({\bf R})  \rangle} \ .
\label{dmc.variat}
\end{equation}

The expectation values obtained through the extrapolation method
(\ref{dmc.extrap}) depend on the
trial wave function $\psi$ used for importance sampling. Therefore,
in spite of using good trial wave functions, the extrapolated
estimator is always biased in a quantity difficult to assess a priori. In
order to overcome that important restriction, one can go a step further
and calculate {\em pure} (exact) expectation values,
\begin{equation}
\langle O({\bf R})
\rangle_p=
\frac{\langle \Phi_0 ({\bf R}) \, | \, O({\bf R}) \, | \,
\Phi_0({\bf R}) \rangle}
{\langle \Phi_0({\bf R}) \, | \, \Phi_0({\bf R})  \rangle} \ .
\label{dmc.pure}
\end{equation}                                          
Having in mind that walkers evolve according to the mixed distribution
$\psi \Phi_0$, the pure estimator is more conveniently written as
\begin{equation}
\langle O({\bf R}) \rangle_p=
\left\langle \Phi_0 ({\bf R}) \, \left| \,
O({\bf R})
\, \frac{\Phi_0({\bf R})}{\psi({\bf R})} \, \right | \,
\psi({\bf R}) \right
\rangle \, \left/ \,
\left\langle \Phi_0 ({\bf R}) \, \left| \,
 \frac{\Phi_0({\bf R})}{\psi({\bf R})} \, \right | \,
 \psi({\bf R}) \right
\rangle  \right.  \ .
\label{dmc.pure2}
\end{equation}              

Some time ago, Liu {\it et al.} \cite{liu} proved that
$\Phi_0({\bf R})/\psi({\bf
R})$ can be obtained from the asymptotic offspring of the ${\bf R}$
walker.  Assigning to each walker ${\bf R}_i$ a weight
$W({\bf R}_i)$ proportional to its number of future descendants
\begin{equation}
W({\bf R})=n({\bf R},t \rightarrow \infty) \ ,
\label{dmc.pes}
\end{equation}
Eq.  (\ref{dmc.pure2}) becomes
\begin{equation}
\langle O({\bf R})
\rangle_p=\frac{ \sum_i O({\bf R}_i) \, W({\bf R}_i)}
{\sum_i W({\bf R}_i)} \ ,
\label{dmc.magia}
\end{equation}
where the summatory $\sum_i$ runs over all walkers
and all times in the asymptotic regime. The difficulty of the method, known
as forward walking, lies on the estimation of the weight $W({\bf R})$
(\ref{dmc.pes}). The weight of a walker existing at time $t$
is not known until a future time  
 $t^{\prime} \geq t+T$,  $T$ being a time interval long enough so
that Eq. (\ref{dmc.pes}) could be replaced by $W({\bf R}(t))=n({\bf
R}(t^{\prime}))$.                

The evaluation of Eq. (\ref{dmc.magia}) has traditionally required the
implementation of a tagging algorithm \cite{reynolds2,barnett1,runge}. The purpose of that algorithm is to
know, at any time during the simulation, which walker of any
precedent configuration originated a present walker. In this way, one can
determine the number of descendants of each ${\bf R}_{i}$, and accumulate its
contribution to Eq. (\ref{dmc.magia}) from the distance. 
An easier method consists in working only with the present values
of $O({\bf R}_{i})$, in such a way that a weight proportional to its future
progeny is automatically introduced \cite{casulleras1}.               
The schedule of this second algorithm is the following. The set of walkers
at a given time $\{ {\bf R}_i \}$, and the values
that the operator $O$ takes on them $\{O_i\}$, evolve after a time step to
$ \{ {\bf R}_i^{\prime} \}$ and $\{ O_i^{\prime} \}$, respectively.
In the same time step, the number of walkers $N$ changes to $N^{\prime}$.
In order to sample the pure estimator of $O$, we introduce an auxiliary variable
$\{P_i\}$, associated to each walker, and that evolves with it (its final average
value will provide the pure expectation value of $O$). The evolution law for $\{P_i\}$ is
given by
\begin{equation}
\{P_i \} \rightarrow \{P_i^{\prime} \} =
\{P_i^t \} + \{ O_i^{\prime} \} \ ,
\label{dmc.patum}
\end{equation}
where $\{P_i^t\}$ is the old set $\{P_i\}$  transported to the
new one, in the sense that each element $P_i$ is replicated as many
times as the ${\bf R}_i$ walker, without any other changes. $\{ P_i
\}$ is initialized to zero when the run starts.

In order to better fulfill the asymptotic condition (\ref{dmc.pes}) the
evolution of the $\{P_i\}$ values is further carried on by means of
\begin{equation}
\{P_i \} \rightarrow \{ P_i^{\prime} \}=\{P_i^t\} \ .
\label{dmc.transport}
\end{equation}
Since a calculation is usually divided in blocks, one can
collect/transport data during a block via Eq. (\ref{dmc.patum}), and allow for a
only reweighting by means of Eq.  (\ref{dmc.transport}) in the following one.
Thus, after a first initialization
block, each new block gives a value for the pure expectation value
of $O$
\begin{equation}
\langle O({\bf R}) \rangle_p = \sum_{i=1}^{N_f} \{P_i\}\, /
\, (M \times
N_f) \ ,
\label{dmc.magic2}
\end{equation}         
$M$ being the length of the block and $N_f$ the number of walkers at the
end of each reweighting block. A  pseudo code for the pure
estimation is available in Ref. \cite{AppendixA}. It is worth mentioning that the algorithm can be
easily generalized to DMC programs that work with branching weights.

The pure estimation depends on the size $M$ of the block. 
$M$ has to be large enough to match the asymptotic regime required by 
the forward walking property (\ref{dmc.pes}).
With an easy
modification of the code reported in Ref. \cite{AppendixA},  one can introduce a pure
estimator for a set of increasing values of $M$ in a single calculation. 
Looking at the results
obtained, as a function of the block size, one observes a characteristic
value $M_c$ from which on the pure estimator is the same (considering the
respective statistical errors).

\section{Results}
\label{results}

The efficiency and accuracy of both the QDMC method and the pure
estimator described in the previous section are analyzed in a calculation
of ground-state properties of atomic Li. 
This is a  well-known system,
with nearly exact results, and serves well to our purpose since it
contains the ingredients that can hinder the pursued second-order behavior: 
a nodal surface and cusp conditions. The present calculation works in the
fixed-node approximation and therefore the energy obtained is an
upper-bound to the ground-state energy of the atom. The trial wave function
used for the importance sampling is the usual Jastrow-Slater form
\begin{equation}
\psi = \psi_{\rm J} \, \phi^{\uparrow} \, \phi^{\downarrow} \ ,
\label{psitrial}
\end{equation}
with $\phi^{\uparrow}$ ($\phi^{\downarrow}$) the Slater determinant of spin
up (spin down) electrons. The two-body Jastrow factor is taken as
\begin{equation}
\psi_{\rm J}= \prod_{i<j} \exp \left( \frac{C_{ij} r_{ij}}{1+b r_{ij}}
\right)   \ ,
\label{psijas}
\end{equation}
with coefficients $C_{ij}=1/2$ ($C_{ij}=1/4$) for spin-unlike (spin-like)
electrons to satisfy the electron-electron cusp conditions.

The functional forms for the orbitals entering in $\phi$ are taken from 
Ref. \cite{langfelder},
\begin{eqnarray}
\chi ^{1s} & = & e^{-\zeta_{1s} r}\, e^{-w r/(1+v r)} \\ \nonumber
\chi ^{2s} & = & (1+cr)\, e^{-\zeta_{2s} r}\, e^{-w r/(1+v r)}  \ .
\label{orbitals}
\end{eqnarray}
They introduce an additional exponential, with respect to hydrogenic-like
orbitals, to better fit the electron-nuclear correlations. The
fulfilment of the electron-nuclear cusp condition introduces two
constraints that reduce the number of parameters to be variationally
optimized to four: $b$, $v$, $\zeta_{1s}$, and $\zeta_{2s}$. We have taken
for them the values reported by Langfelder {\em et al.} \cite{langfelder}. 

The VMC energy obtained with $\psi$ is in agreement with the one reported
in Ref. \cite{langfelder}, $E=-7.4737(1)$. That supposes a 90\% of the
correlation energy, which is significantly high taking into account the
relatively simple trial wave function used. That result coincides with the more
sophisticated wave function $\psi_7$ proposed by Schmidt and Moskowitz
\cite{mosko}, and is slightly worse than the energy obtained using
$\psi_{17}$ ($E=-7.47669$) from the same authors. In a DMC calculation,
like the one intended here, the proposal of Langfelder {\em et al.} 
\cite{langfelder} is more appropriate since it satisfies both the
electron-electron and electron-nuclear cusp conditions. These requirements
are essential in order to eliminate divergences of the drift force that
break down the expected time-step accuracy. In fact, we have verified that if
$\psi_7$ is used instead of $\psi$ a second-order behavior is not achieved. 

One of the main concerns of the present work is the time-step dependence of
the different DMC algorithms discussed in Sec. II. In Table I, 
results for the energy as a function of $\Delta t$ are reported using
linear DMC, linear DMC with rejection (MDMC) and second-order DMC (QDMC).  
The {\em exact} non relativistic ground state energy is  
-7.478 060 323 10(31), calculated variationally by using a multiple basis
set in Hylleraas coordinates \cite{yan}. The last row in Table I contains
the extrapolation to $\Delta t =0$ based on a polynomial fit to the data.
The QDMC results follow the expected second-order behavior 
\begin{equation}
E(\Delta t)= E_0 +E_2 ~ (\Delta t)^2
\label{anal2}
\end{equation}
with good accuracy ($\chi^2/ \nu = 1.2$). The extrapolated energy
$E_0=-7.47805(2)$, which reproduces the exact value, is statistically
indistinguishable from the energy obtained with $\Delta t=0.003$. In fact,
one may move in a range $\Delta t=0.0025-0.0035$ without significant
changes in the energy. This is a very useful feature which derives from the 
second-order Green's function used in the QDMC method, and that has been
also observed in quantum liquids calculations \cite{boro4he,boro3he}.   

The extrapolations corresponding to the DMC and MDMC calculations have been
carried out with a second-degree polynomial that, 
differently from the QDMC
case (\ref{anal2}), includes a linear term. The variance of the DMC result 
is larger than the QDMC one and the energy obtained is slightly above 
the QDMC and exact energies. The different dependence on $\Delta t$ of the
DMC and QDMC algorithms is illustrated in Fig. 1a: DMC energies are
essentially linear while QDMC ones are quadratic.   

The introduction of an accept/reject step in the imaginary-time evolution
modifies completely the time-step dependence of the DMC method. As is clear
from Table I and Fig. 1b the MDMC method allows for a significant reduction
of the time-step bias, mainly for large $\Delta t$. For values $\Delta t
\geq 0.01$, the MDMC energies are  closer to the exact result than 
the QDMC ones.  However, when
$\Delta t$ is reduced the behavior of the energy shows a quite erratic
behavior (see Fig. 1c) that makes difficult the extrapolation to zero 
time-step. The extrapolated value reported in Table I comes from a
second-order polynomial fit. The value obtained, -7.47810(14) is in
agreement with the QDMC result but its variance is an order of magnitude
larger.

The pure estimator described in Sec. II, in conjunction with the QDMC
method, has been used to calculate expectation values of several radial
operators. In particular, we have considered  
the first radial moments of both the one-body density and
the two-body intracule and extracule densities \cite{coleman}, defined as
\begin{eqnarray}
\langle r^n\rangle&=&\frac{\langle \Psi|\sum\limits_{i=1}^N r^n |\Psi\rangle}
             {\langle \Psi|\Psi\rangle} \nonumber \\
\langle r_{ij}^n\rangle&=
&\frac{\langle \Psi|\sum\limits_{i<j=1}^N r_{ij}^n |\Psi\rangle}
                  {\langle \Psi|\Psi\rangle} \nonumber \\
\langle R^n\rangle&=&\frac{\langle \Psi|\sum\limits_{i<j=1}^N R^n |\Psi\rangle}
             {\langle \Psi|\Psi\rangle}    \ . 
\label{defmom}
\end{eqnarray}
In Eqs. (\ref{defmom}),  ${\bf r}_i$ is the position vector of the $i$-electron, 
${\bf r}_{ij}={\bf r}_i-{\bf r}_j$, and 
${\bf R}=({\bf r}_i+{\bf r}_j)/2$ is the center of mass
vector of  electrons $i$ and $j$. Those observables play an important
role in the description of the structure and
dynamics of atoms and molecules 
\cite{coleman,thakkar}. They have been calculated 
within different theoretical frameworks 
\cite{king1,yan,king2,galvez1,coleman,thakkar,esquivel1,esquivel2,barnett2,alexander,meyer,koga1,koga2,sarsa2,galvez2,worsnop,fradera}
to determine electronic properties, to interpret the Hund rules, 
to elucidate electron correlation effects,
or to study elastic and inelastic form factors.  

In Fig \ref{fig2}, we
show the dependence of the pure expectation values of $\langle r\rangle$, 
$\langle r_{12}\rangle$, and $\langle R\rangle$ upon the number of time steps $M$ of a block. 
In these calculations, more CPU time has been
invested for the largest values of the block length $M$ in order to reduce 
the statistical noise. Nevertheless,  
the pure estimators reach very soon the asymptotic regime,  for $M$ values 
as low as $M \simeq 500$. The statistical error of the pure
estimators, evaluated in the asymptotic regime, is typically less than twice
the statistical error of the mixed estimators for a fixed CPU time.
The asymptotic results, i.e., the {\em real} pure estimators,
reproduce accurately the exact values. As a matter of comparison, the mixed
estimations and the variational ones are also
shown. With the mixed and variational estimators, one easily obtains
the extrapolated values (\ref{dmc.extrap}) that are also plotted in Fig. 2.
In all cases, the extrapolated estimator does not match the exact values, 
manifesting a bias that depends on the particular operator. 
The above features are also applicable to the second-degree radial
operators $\langle r^2\rangle$, $\langle r_{12}^2\rangle$, and $\langle R^2\rangle$ which appear reported in Fig.
3. 

Our best pure estimates for the $\langle r^n\rangle$, $\langle r_{12}^n\rangle$, and $\langle R^n\rangle$ moments
are contained in Tables II, III, and IV, respectively. Within the
statistical errors, all the pure results are in agreement with the values
considered as exact. With nearly the same variance than the extrapolated
values,  pure estimators
eliminate the bias which comes from the trial wave function. On the other
hand, extrapolated values retain some bias from $\psi$ which is 
certainly difficult to assert a priori. It is worthwhile determining the
accuracy of pure estimators using different propagators like DMC or MDMC. 
For that purpose, we have also implemented the pure estimator algorithm to
the MDMC program. The results obtained are reported in Tables II-IV. The
pure MDMC results improve the extrapolated values but do not share the
excellent quality obtained with the QDMC algorithm. Somehow, the reason for 
that is the use of effective time-steps in the branching factor, the 
accuracy of which is a crucial point to ensure the success of the method.

\section{Conclusions}
\label{conclusions}

In the present work, the time-step dependence of the diffusion Monte Carlo
method and the implementation of pure estimators, both in atomic systems,
have been addressed. The study has been carried out for Li since it already
contains the necessary ingredients to test the method and, additionally, 
a lot of nearly exact data are available. Using a
 short-time Green's function accurate to  ${\cal O}(\Delta t)^3$, we
 have proved that it is possible in atomic systems to have a quadratic
 behavior in the simulated energy as was reported in quantum-liquid 
 calculations. To the best
 of our knowledge it is the first stringent test of QDMC on atomic systems
 with nodal surface. The accuracy obtained had been occasionally questioned
 due to the presence of the divergences that appear in the drift force near
 the nodes. The present results show that the presence of nodes is not a 
 real handicap, a conclusion that coincides with the one achieved in  a 
 recent QDMC calculation of homogeneous (fermion) liquid $^3$He
 \cite{boro3he}. The most
 crucial point to guarantee a success of the QDMC method is the correct
 fulfilment of the electron-electron and electron-nuclear cusps conditions.
 The energies obtained with the relatively simple model for $\psi$, proposed
 by Langfelder {\em et al.} \cite{langfelder}, follow accurately the
 expected second-order behavior. Additional checks performed using better
 trial wave functions, from the variational view, but not satisfying the
 electron-nuclear cusp  show a more erratic behavior of the function
 $E(\Delta t)$. 

The second aim of the work has been the consideration of a rather simple pure
estimator, based on the forward walking methodology. Pure results for
different radial moments are in agreement with exact values and are
manifestly superior to the extrapolated estimations.

As a concluding remark, we think that the combination of a QDMC method and
the pure estimation here presented constitutes an appropriate tool for studying
atoms and small molecules. Moreover, and very importantly from the
programmer's viewpoint, the effort {\it i}) moving from a usual DMC to QDMC, and
{\it ii}) introducing a pure estimator, is really small.

\acknowledgments 
This research has been partially supported by DGESIC (Spain) Grant N$^0$
PB98-0922, and DGR (Catalunya) Grant N$^0$
1999SGR-00146. A.S. acknowledges the Italian MURST for financial support.

\begin{table}
\caption{Energy of the Li atom versus the time step and using  
different algorithms. The figures enclosed in parenthesis are the 
statistical errors. The {\em exact} energy is -7.4780632310(31) 
}
\label{tab1}
\begin{tabular}{dddd}
$\Delta t$ & DMC      & MDMC        & QDMC \\
\tableline
0.030         & -7.45279(4) & -7.47860(4) & -7.48244(3) \\
0.0200        & -7.46074(5) & -7.47844(3) & -7.48007(3) \\
0.0100        & -7.46893(5) & -7.47831(4) & -7.47855(4) \\
0.0075        & -7.47127(4) & -7.47810(3) & -7.47831(3) \\
0.0050        & -7.47333(3) & -7.47811(3) & -7.47817(2) \\
0.0030        & -7.47520(4) & -7.47827(4) & -7.47807(4) \\
extrapolation & -7.47784(11) & -7.47810(14) & -7.47805(2) \\
\end{tabular}
\end{table}

\pagebreak

\begin{table}
\caption{Estimations  of  $\langle r^n\rangle$  moments 
of the Li atom using the MDMC and QDMC algorithms. 
 }
\label{tab2}
\begin{tabular}{dddd}
                & $\langle r\rangle$      
&  $\langle r^2\rangle$      & $\langle r^3\rangle$ \\
\tableline
Variational\tablenotemark[1]              & 4.9553(7)  
                         & 18.088(7)      & 91.12(6) \\ 
Extrapolated (MDMC)      & 4.992(3)  
                         & 18.39(3)                   & 92.8(3) \\
Extrapolated (QDMC)      & 5.001(3)  
                         & 18.46(3)                   & 93.3(3) \\
Pure (MDMC)              & 4.985(2)  
                         & 18.32(2)                   & 92.4(2) \\
Pure (QDMC)\tablenotemark[2]               & 4.991(2)  
                         & 18.36(2)                   & 92.7(2) \\
{\em exact}              & 4.989523\tablenotemark[3]    
                         & 18.354615\tablenotemark[3] 
                         & 92.60364\tablenotemark[4] \\
\end{tabular}
\tablenotetext[1]{VMC results of Ref. \cite{langfelder}: 
$\langle r\rangle = 4.9536(6)$, $\langle r^2\rangle = 18.076(6)$}
\tablenotetext[2]{GSD-DMC results of Ref. \cite{langfelder}:
$\langle r\rangle =4.992(7)$, $\langle r^2\rangle =18.36(7)$}
\tablenotetext[3]{Hylleraas expansion of Ref. \cite{yan}}
\tablenotetext[4]{Hylleraas expansion of Ref. \cite{king1}}
\end{table}

\pagebreak

\begin{table}
\caption{Estimations  of $\langle r_{12}^n\rangle$  moments 
of the Li atom using the MDMC and QDMC algorithms.}

\label{tab3}
\begin{tabular}{dddd}
                        & $\langle r_{12}\rangle$     
                        &  $\langle r_{12}^2\rangle$ 
                        & $\langle r_{12}^3\rangle$ \\
\tableline
Variational             & 8.593(2)
                        & 36.32(1)                   & 189.2(1) \\ 
Extrapolated (MDMC)     & 8.673(7)  
                        & 36.89(6)                   & 192.4(5) \\
Extrapolated (QDMC)     & 8.686(7)  
                        & 37.02(6)                   & 193.5(5) \\
Pure (MDMC)             & 8.662(3)  
                        & 36.79(3)                   & 191.6(3) \\
Pure (QDMC)             & 8.671(3)  
                        & 36.87(3)                   & 192.4(3) \\
{\em exact}             & 8.668397\tablenotemark[1]    
                        & 36.847838\tablenotemark[1] 
& 192.10037 \tablenotemark[2]  \\
\end{tabular}
\tablenotetext[1]{Hylleraas expansion of Ref. \cite{yan}}
\tablenotetext[2]{Hylleraas expansion of Ref. \cite{king3}}
\end{table}

\pagebreak

\begin{table}
\caption{Estimations  of $\langle R^n\rangle$  moments 
of the Li atom using the MDMC and QDMC algorithms.}

\label{tab4}
\begin{tabular}{dddd}
              & $\langle R\rangle$      
&  $\langle R^2\rangle$      & $\langle R^3\rangle$ \\
\tableline
Variational              & 4.2615(7)
                         & 9.009(3)                 & 23.48(2) \\ 
Extrapolated (MDMC)      & 4.301(3)  
                         & 9.15(2)                  & 23.9(4) \\
Extrapolated (QDMC)      & 4.309(4)  
                         & 9.19(2)                  & 24.05(7) \\
Pure (MDMC)              & 4.309(3)  
                         & 9.19(2)                  & 23.76(7) \\
Pure (QDMC)              & 4.300(2)  
                         & 9.150(8)                 & 23.89(4) \\
{\em exact}              & 4.2996(6)\tablenotemark[1]    
                         & 9.145(3)\tablenotemark[1]  
                         & 23.86(1)\tablenotemark[1]  \\
\end{tabular}
\tablenotetext[1]{Ref.\cite{galvez1}}
\end{table}

\begin{figure}[t]
\caption{
Total energy of the Li atom versus the  time step $\Delta t$ computed by
using the DMC, MDMC and QDMC algorithms. The lines on top of the data
correspond to polynomial fits: 
$E(\Delta t)=E_0+E_1 \Delta t+E_2 (\Delta t)^2$ for 
DMC and MDMC, and $E(\Delta t)=E_0+E_2 (\Delta t)^2$ for QDMC. 
In all cases, the error bars are smaller than the size of the symbols. The
dotted line stands for the exact energy. 
In a) the time-step dependence of the DMC and QDMC energies are
compared. In b) the comparison is between QDMC and MDMC. Finally, c)
illustrates the different behavior of MDMC and QDMC when $\Delta t
\rightarrow 0$; notice the  dispersion of the MDMC data compared with the
regular behavior of the QDMC energies. 
}
\begin{center}
\epsfxsize=22pc
\epsfbox{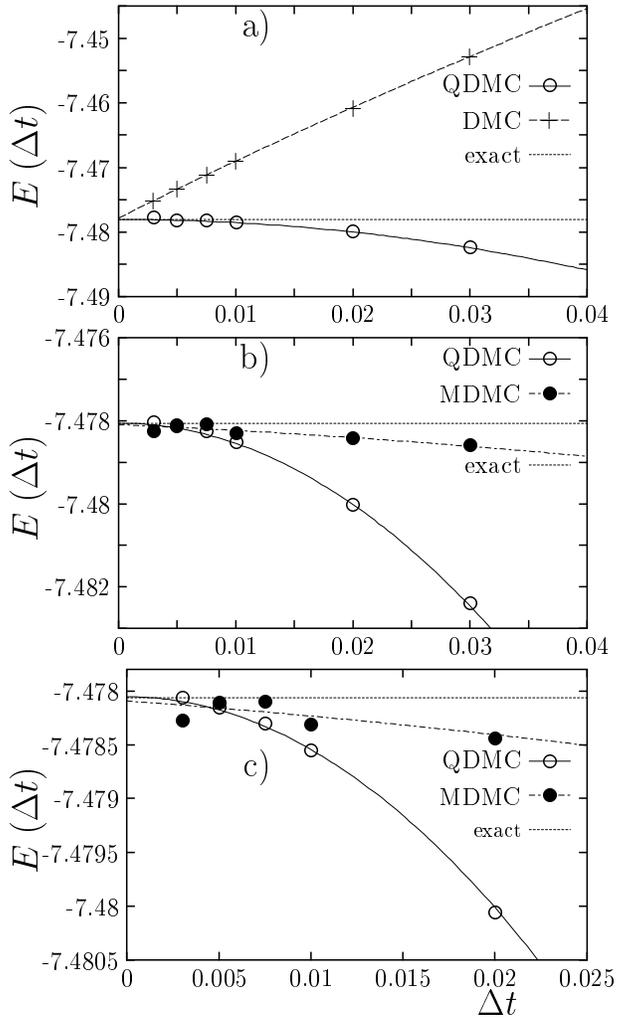}
\end{center}
\label{fig1}
\end{figure}

\newpage

\begin{figure}[t]
\caption{
Pure expectation values of first-degree radial moments of the Li atom
as a function of the block length $M$. 
The solid, dashed, and dotted lines stand for the exact, variational, and 
extrapolated results, respectively.
}
\begin{center}
\epsfxsize=22pc
\epsfbox{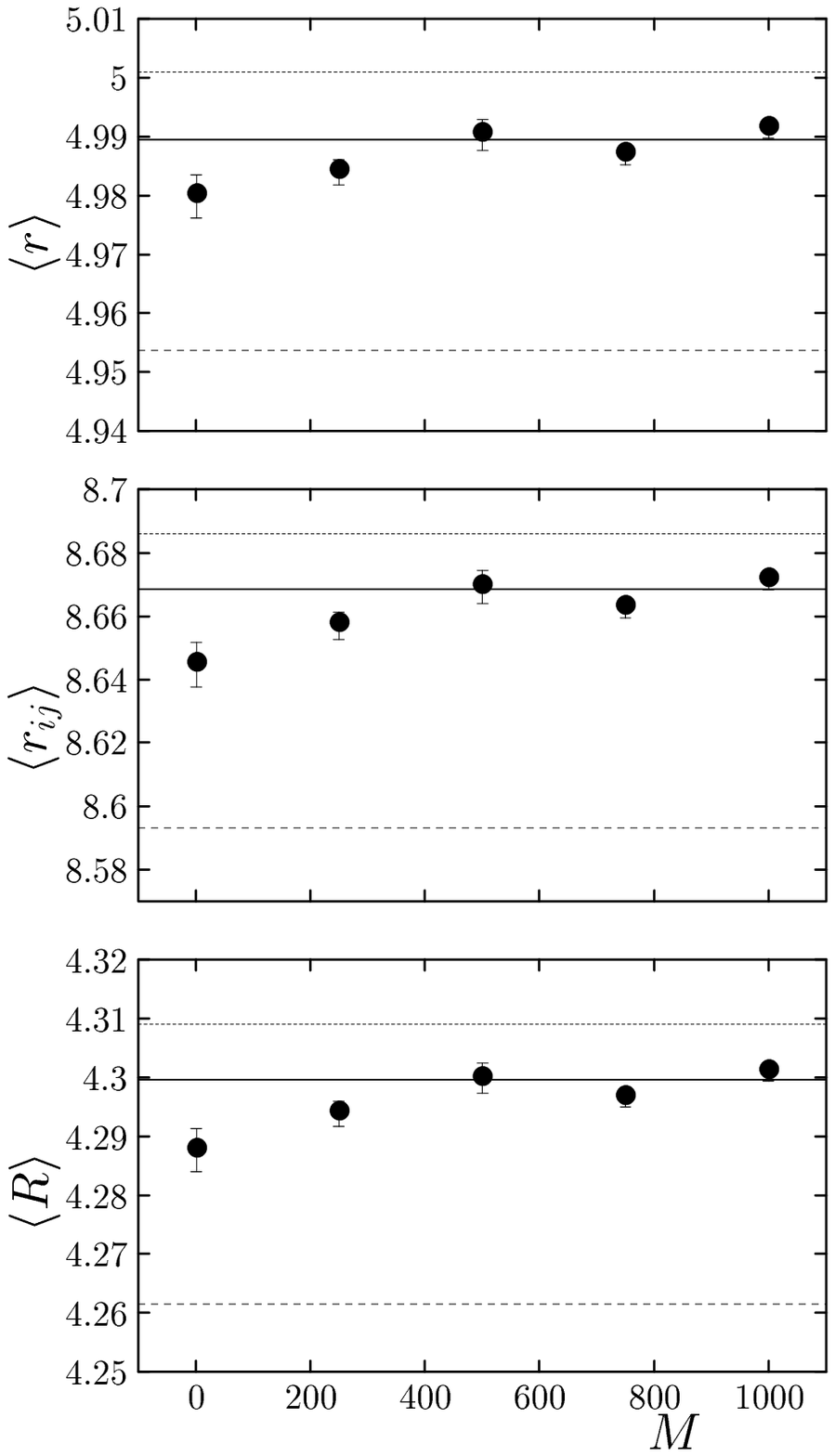}
\end{center}
\label{fig2}
\end{figure}

\newpage

\begin{figure}[t]
\caption{
Pure expectation values of second-degree radial moments of the Li atom
as a function of the block length $M$. 
The solid, dashed, and dotted lines stand for the exact, variational, and 
 extrapolated results, respectively.
}
\begin{center}
\epsfxsize=22pc
\epsfbox{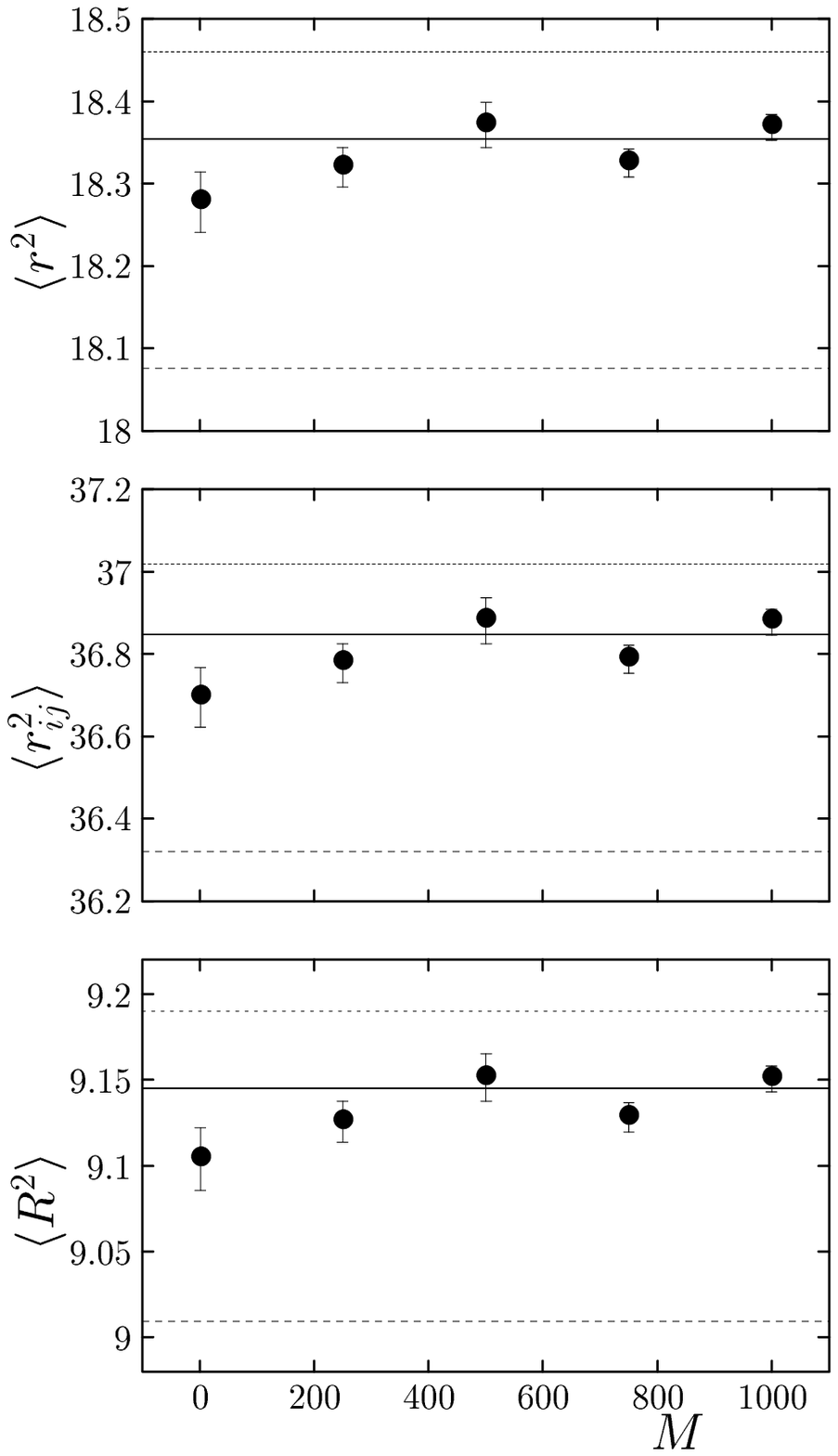}
\end{center}
\label{fig3}
\end{figure}

\end{document}